# Can the expansion of the universe localize quantum wave functions? How classical behavior may result from Hubble expansion

C. L. Herzenberg

**Abstract**
We consider an object at rest in space with a universal Hubble expansion taking place away from it. We find that a governing differential equation developed from the Schrödinger equation leads to wave functions which turn out to exhibit pronounced central localization. The extent of concentration of probability depends on the mass; objects with small masses tend to behave in a delocalized manner as ordinary quantum objects do in a static space, while quantum objects with large masses have wave functions that are largely concentrated into much smaller regions. This in turn suggests the possibility that classical behavior is being induced in quantum objects by the presence of the Hubble expansion. If the size of the localized region of concentrated probability density is larger than the size of the corresponding extended object, quantum behavior might be expected; whereas classical behavior might be expected for cases in which the region of high probability density is smaller than the size of the object. The resulting quantum-classical boundary due to Hubble expansion may be expressed in terms of a relationship between the size and mass of an object, or may be expressed in terms of a threshold moment of inertia.

**Introduction**

How will quantum objects behave in an expanding universe? This is a challenging, deep, and difficult question. However, it can sometimes be useful to attempt to elucidate even very complicated and difficult problems by initially using simple and straightforward approaches. That is what will be attempted here.

We will approach this question using elementary quantum mechanics in an attempt to examine what features might be expected for the quantum behavior of an object at rest in an expanding universe, and what the implications of these features may be.

**Characteristics of cosmological expansion**

Cosmological expansion, or Hubble expansion, is widely regarded as originating from the expansion of the universe, and to be a characteristic of space itself. The metric expansion of space is usually considered an intrinsic expansion and is a key feature of the Big Bang cosmology.[1]

The existence of Hubble expansion on cosmological scales has been very well established for many years. The velocity of expansion is observed to be radially outward away from the point of observation and to increase proportionately with the distance; this can be expressed quantitatively as:

$$v(r) = H_o r \qquad (1)$$

Here, r is the radial distance away from a defined position which we will take as the origin of a coordinate system, v is the radial expansion velocity, and $H_o$ is the constant of proportionality, the Hubble constant.

We will use an idealized model of an expanding universe, in which the expansion is uniform and is treated as taking place at every point in the universe.

We will be interested in examining the behavior of quantum wave functions for describing the behavior of quantum objects in such an expanding space. We will be concerned with the behavior of wave functions largely in the vicinity of the quantum objects themselves, and we will not treat wave function behavior at extremely large distances where relativistic expansion velocities would be involved.

A more detailed discussion of cosmological expansion in this context may be found in Appendix A.

**An approach to characterizing quantum wave behavior in an expanding universe**

Adequately addressing the problem of developing quantum wavefunctions to characterize objects in an expanding universe would demand the use of advanced techniques and is far beyond the scope of this study. However, it would seem to be possible to initiate a more simplified approach and make some general observations and obtain some estimates that might at least approximate some aspects of quantum wave behavior.

We are faced with the problem of describing the presence of expansion of the space in which the quantum object resides. Phenomenologically, observations tell us that distant positions in space, or at least the objects located at them, recede at rates proportional to their distance from the observer. Can we base a quantum description on this information?

How might we regard a wave function in an expanding universe? Observations only indicate that different objects recede with respect to each other, but would an object recede with respect to itself in its probability distribution description? Should quantum wave behavior only describe an object's distribution with respect to the space which is itself expanding, or could a wave description include the behavior of the expanding space? Expansion is usually described with respect to a single point of observation, which leads to a unique coordinate system; could the description of a proposed quantum wave with respect to such a unique coordinate system be of any value?

While the universe as a whole is expanding, our part of the universe seems reasonably quiescent, so we might seek quantum wave behavior that is quasi-steady-state, at least nearby.

First, we will take a qualitative look at what we might expect for behavior of such wave functions, and then make some estimates of how far such a wave function might extend within an expanding space. Then we will formalize the discussion by introducing a proposed governing equation based on the Schrödinger equation, and then we will examine solutions of the governing equation. Following this, we will examine the implications of these results for the quantum-classical transition, and for quantum and classical behavior in general.

Parenthetically, it should be noted that the size estimates and numerical results presented in this paper are generally semiquantitative, good to an order of magnitude at best.

**A quick introductory look at the concentration of quantum probability density and the resultant localization of quantum objects in an expanding space**

First, we will engage in a quick overview of some qualitative and semiquantitative aspects of quantum behavior that might be expected to occur for quantum wave functions in the presence of Hubble expansion, to be followed in succeeding sections by a more careful look at these same issues.

We consider an object at rest in space with Hubble expansion taking place away from it. We propose a trial wave function that is nearly constant (corresponding to a near-infinite wavelength) at the center so as to describe at that location an object in a state of rest in the absence of outward flow. To describe a symmetric outward flow from a central location, we would need a radially symmetric wave function describing a flow. Constant flow would correspond to a radially symmetric wave function dropping off as $1/r$, and incorporating an exponential propagating wave form $e^{2\pi i(kr-ft)}$ describing the flow. (Henceforward we will omit the temporal dependence of wave functions.) While Hubble flow is not a situation of constant flow it would appear that it might be described by a quasistatic flow. The wave function would then be expected to decrease in amplitude as it expands into three dimensional space, and also as it represents a faster flow at increasing radial distances. We consider fitting a trial wave function with a wavelength appropriate to the recession velocity as a function of distance. Since the wavelength associated with any location would be expected to decreases as the velocity increases, the wavelength would become progressively shorter further away from the origin, and the corresponding amplitude would be expected to drop off so as to accommodate to a quasi-steady-state flow. At any location, the wave could be expected to turn over appreciably in about a quarter wavelength of oscillation. Accordingly, where the wavelength becomes comparable to the distance from the origin, we can expect that will establish a rough size for the central region of high amplitude of the wave. Thus, using the criterion $r \approx \lambda/4$, we can estimate the size of the region of high probability density. But since $\lambda = h/mv$, and $v = H_o r$, we find $\lambda = h/mH_o r$ and then we find for an estimate of the radius of the region of large probability density for the quantum wave that $r \approx \lambda/4 \approx h/4mH_o r$, and hence the radial extent of the central region of high probability density $r_c$ can be estimated approximately by:

$$r_c^2 \approx h/4mH_o \qquad (2)$$

This gives us an estimate of the radius of the central region where the wave function is expected to be large, and hence of the region of high probability density for finding the quantum object. Because of the inverse dependence on mass, it is clear that this region where the wave function is concentrated would be smaller for objects with higher masses.

Now, we will examine the approach that we have just sketched out in a somewhat more careful manner and in somewhat more detail.

**Qualitative and semiquantitative considerations regarding wave function behavior in an expanding space**

Let us consider ab initio how the wave associated with a quantum particle at rest in an expanding universe might be expected to behave. In the Hubble expansion, the velocity is zero at the origin of coordinates, and increases with increasing radial distance away. Therefore, the wave function might be expected to describe a state of rest at the origin of coordinates; while at any distance away from the origin, the wave function would seem to need to describe a quantum wave in motion radially away from the origin, and exhibiting a speed of recession that increases linearly as a function of distance from the origin.

We will approach this problem by looking initially at more familiar situations in ordinary quantum mechanics in a static universe.

**Static and flow behavior in ordinary elementary quantum mechanics in a fixed Euclidean space for comparison purposes**

We will first take note of how a quantum object at rest is described in ordinary quantum mechanics in the absence of spatial expansion. Under these circumstances, the quantum wave function for an object at rest would be uniform and identical throughout all of space, being essentially a constant amplitude standing wave of infinite wavelength.

However, if instead we wanted to describe a free quantum object in a state of motion diverging radially away from the origin of the coordinate system (in the absence of spatial expansion), we would express the time-independent Schrödinger equation in polar coordinates.[2] If there is no potential and no angular momentum present, a spherically symmetric outgoing wave function would describe the radial flow. A solution would have the form $(1/r)e^{2\pi i k r}$, where k is the wave number and is given by $k = 1/\lambda$, and where the wavelength $\lambda = h/mv$.[2] This solution corresponds to a constant wave number and constant wave length and provides for continuity with conservation of probability as a function of r so that the net flow is the same through all spherical shells. The solution has a source at the origin of coordinates and is steady-state, so that it does not deplete the probability density at any location with time. This solution is infinite at the origin of coordinates and finite elsewhere, and exhibits wave structure throughout space, with the magnitude of the wave decreasing with distance from the origin.

It is interesting to observe that a wave function describing a free particle at rest with no flow is completely delocalized throughout the universe, showing no region of concentration anywhere; whereas, a wave function describing a free particle flowing radially outwards describes a free particle that is overwhelmingly localized and concentrated near the origin of flow. These qualitative features already present in ordinary elementary quantum mechanics seems to suggest that outward flow of a quantum wave might contribute to the localization of the objects that we observe in our classical world. Thus, basic results in elementary quantum mechanics already seem to be suggesting to us that radial flow can result in concentration and localization, and hence might be involved in bringing about a type of quantum behavior resembling classical behavior.

**Characterizing a quantum wave in an expanding universe: Bringing in Hubble expansion**

We have compared two cases in elementary quantum mechanics: A quantum object at rest with no flow is spread out uniformly throughout all space; while a quantum object flowing out radially in three dimensional space exhibits a probability density distribution that is infinite at the origin and drops off fairly rapidly in the radial direction. We can use these cases to try to help our understanding of quantum behavior in an expanding universe. Since there is no flow at the origin and an increasing flow at increasing radial distances outwards in the description of an expanding universe, an intermediate situation might be expected. One might speculate that the quantum wave magnitude would be large but finite and relatively uniform in the vicinity of the origin, while decreasing in magnitude with distance from the origin.

Another intermediate case would be presented by a spherically symmetric wave packet centered on the origin of coordinates that would have a finite magnitude near the origin and drop off to small values at large distances away; however, such a wave packet would not represent a steady-state solution for a freely moving object in a static Euclidean space.

If we are dealing with an expanding universe that is intrinsically changing with time, we cannot expect time-independent behavior for quantum waves. However, at least over periods of time short compared to the lifetime of the universe, behavior that is slowly varying in time might be anticipated.

We would want to attempt to describe a quantum object that is not undergoing major outflow at its central location, the origin of coordinates, but undergoes increasing speeds of flow away from the origin at increasing distances away from the origin. The flow would be expected to reflect effects associated with both the falloff of probability density with distance that occurs in constant flow as well as the additional decreasing probability density associated with the increase of flow velocity with distance to be expected under quasi-steady-state conditions.

We can presumably expect the highest probability of finding the quantum object right at the position where the associated classical object would be located, at r = 0. Thus, we might anticipate the largest magnitude for the wave function at r = 0. The quantum wave function would be expected to exhibit motion radially outward in association with the radial expansion velocity. Thus, we might expect outgoing waves with locally determined wavelengths that shorten as a function of the distance away from the origin. The wavelength as a function of radial distance would be expected to be given approximately by λ(r) = h/p(r) = h/mv(r).

The flow is outward, so the probability of finding the object at the origin would be expected to decrease with time as the universe expands. If we are dealing with a quasi-steady-state flow, the probability of finding the object at other locations would also decrease with time comparably except that the wave function would be expected to extend to greater and greater distances as the universe expands. Since the flow would be expected to vary roughly with the velocity, which increases radially, we might expect the probability of finding the object at any radial distance to decrease with distance, to accommodate a quasi-steady-state flow.

Thus, we can expect approximations that describe a wave function that has a magnitude that drops off fairly rapidly with radial distance at least for sufficiently large values of the mass m, and exhibits periodic behavior with a wavelength that grows progressively shorter as the radial distance from the origin increases.

**Estimating wave function behavior near the origin of coordinates**

This description would seem to correspond roughly to a wave function that behaves near the origin in a nearly uniform manner but exhibits some wave structure, while at greater distances from the origin it would continue with increasingly short wavelength oscillations but also drop off fairly rapidly with the distance at intermediate to larger distances.

Thus, near the origin, we might expect a spatial behavior similar to:

$$\psi(r) = A\cos(2\pi k(r)r) = A\cos(2\pi r/\lambda(r)) = A\cos(2\pi r^2 mH_o/h) \qquad (3)$$

Here, k(r) is the wave number or inverse of the wavelength, $k = 1/\lambda(r) = mv(r)/h = mH_o r/h$, and A is a multiplicative factor that would be expected to be close to being constant near the origin of coordinates. This meets the requirements of providing a nearly constant probability density near the origin, while exhibiting a wave structure having increasingly short wavelength as a function of distance from the origin.

Further out, we can expect to describe a wave function that has a magnitude that drops off fairly rapidly with radial distance at least for sufficiently large values of the mass m, and exhibits periodic behavior with a wavelength that grows progressively shorter as the radial distance from the origin increases.

For present purposes, we are mainly interested in the behavior of the wave function at very small values of the radial distance r, in the vicinity of where the corresponding classical object would be located. Equation (3) that would seem to approximate the wave function for small radial distances would be applicable. The cosine function will reach its first zero at one quarter wavelength away from the origin; in this equation that will occur at a radial distance:

$$r_c = (h/4mH_o)^{½} \qquad (4)$$

The similarity of this result to our first estimate in Eqn. (2) can be noted. It would appear that we can use these results to provide a measure for estimating the extent of the quantum wave, since even as the sinusoid starts to recover, the magnitude of the wave can be expected to drop off at larger values of the radial distance in a more realistic estimate that takes into account the radial expansion of the flow into space and the radially increasing speed of flow.

This result appears to be telling us that an object in an expanding universe will be characterized by a localized wave function with an estimated characteristic size given roughly by Eqn. (4). Thus, it appears that an object of mass m in an expanding universe may be characterized by a localized region of high amplitude of the quantum wave function of size roughly given by $(h/4mH_o)^{½}$.

After this qualitative introductory study to assess the situation, we will next approach the problem more systematically by developing a proposed governing differential equation applicable to this situation.

**Developing a governing equation for wave functions in an expanding universe**

How might we go about developing a governing wave equation that could describe the behavior of a quantum particle in an expanding universe? As noted earlier, we will not even attempt to use a relativistic wave equation but simply explore how the non-relativistic Schrödinger equation might be adapted to this problem in the region of non-relativistic behavior. Since we are seeking quasistatic solutions, approximate solutions might be sought based on the use of the time-independent Schrödinger equation. Neither the Schrödinger equation for a free particle nor the usual Schrödinger equation for a particle in a potential would seem to be appropriate in this context. However, the Schrödinger equation for a particle in a potential, which involves the difference between the total energy and the potential energy, can be adapted by being reexpressed in terms of the kinetic energy and thus in terms of the square of the velocity.

In three dimensions, under spherically symmetric conditions, the time-independent Schrödinger wave equation can be written as the product of radial and angular wave functions, with the radial wave equation being given by:[2]

$$(1/r^2)(d/dr)(r^2 d\psi(r)/dr) + (8\pi^2 m/h^2)[E - V(r) - (h^2/8\pi^2 m)\{\ell(\ell+1)/r^2\}]\psi(r) = 0 \quad (5)$$

Here, $\psi(r)$ is the radial wave function, m is the mass, E is the total energy, $V(r)$ is the potential energy, $\ell$ is the angular momentum quantum number for the angular wave function, and h is Planck's constant.

If we limit our attention to the case of wave functions with zero orbital angular momentum, the last term drops out. Under these conditions, the preceding equation would become:

$$(1/r^2)(d/dr)(r^2 d\psi(r)/dr) + (8\pi^2 m/h^2)[E - V(r)]\psi(r) = 0 \quad (6)$$

We note that the situation that we are dealing with is not one of constant total energy, as we are seeking to describe quantum behavior that will correspond to higher velocities and hence higher kinetic energies at larger distances. However, this equation can be reexpressed in terms of velocities, since the difference between the total energy and the potential energy is the kinetic energy K, which is given by $E - V = K = \frac{1}{2}mv^2$. Accordingly, we will reexpress the governing Schrödinger equation in terms of a local velocity function defined by:

$$\tfrac{1}{2} m v(r)^2 = E - V(r) \quad (7)$$

Combining Eqn. (7) with Eqn. (6), we can reexpress the radial Schrödinger equation in terms of the square of the local velocity:

$$(1/r^2)(d/dr)(r^2 d\psi(r)/dr) + (4\pi^2 m^2/h^2)\,v(r)^2 \psi(r) = 0 \quad (8)$$

But, in this particular case the only velocity is an expansion velocity proportional to the radial distance from the origin, in accordance with Eqn. (1).

When we insert Eqn. (1) into Eqn. (8), we obtain the following proposed governing equation:

$$(1/r^2)(d/dr)(r^2 d\psi(r)/dr) + (2\pi m H_o/h)^2 r^2 \psi(r) = 0 \quad (9)$$

Let us pause to examine some features of this governing equation. Bear in mind that this equation would not be applicable for very large values of the radial distance r, since at very large radial distances the Hubble velocities become relativistic, and this governing equation is based on the non-relativistic Schrödinger equation. Therefore, the features of this governing equation will be valid only at values of r that are small compared to the size of the universe.

Let us examine the properties of this governing equation subject to a restriction to small values of r.

First off, if the mass is vanishingly small, the second term in Eqn. (9) will become vanishingly small and the first term will dominate, and, in the limit, we will be left with the first term only. Thus, we will be left with the ordinary Schrödinger equation for a free particle with zero kinetic energy, that is, an ordinary free particle at rest. So, in the limit of sufficiently small masses, objects governed by this equation would behave as ordinary quantum objects at rest in a static Euclidean space.

Secondly, let's look at the limiting case when the mass m in the governing equation is allowed to rise to very large values. Then the second term can remain finite only if the product $r^2\psi(r)$ goes to zero. This will take place if r is zero and/or if $\psi(r)$ is zero. Thus, in the limit of very large masses, it would seem that only near the origin of coordinates (or in the limit, at the origin of coordinates) can the wave function be non-zero. This result therefore suggests that it is an intrinsic feature of this governing equation for quantum objects in an expanding space that it will cause objects of sufficiently large mass to become strongly localized near the origin of coordinates.

Thus, it appears that this governing equation will cause massive objects to be localized, while objects with very small masses will behave as delocalized quantum objects. This result would seem to be relevant to any discussion of quantum versus classical behavior and the boundaries between quantum and classical behavior in our world.

Further discussion of this governing equation for quantum wave functions in an expanding universe, and of its solutions and their properties is to be found in Appendix B.

**Solutions to the governing equation for quantum objects in an expanding space**

As discussed in Appendix B, the general solution to the governing equation for a quantum object whose classical counterpart is at rest in an expanding space, Eqn. (9), can be found in the form:

$$\psi = r^{-\frac{1}{2}} [C_1 J_{\frac{1}{4}} (\pi m H_0 r^2/h) + C_2 Y_{\frac{1}{4}} (\pi m H_0 r^2/h)] \qquad (10)$$

Here $C_1$ and $C_2$ are arbitrary constants. $J_{\frac{1}{4}}$ is a Bessel function of the first kind of fractional order ¼ and $Y_{\frac{1}{4}}$ is a Bessel function of the second kind (also called a Neumann function) of fractional order ¼.

Both terms are oscillatory, with the amplitudes decreasing as the radial distance increases. The first term has its largest value near the origin but remains finite, while the second term becomes negatively infinite at the origin.

While the second term of the general solution has a singularity at the origin, this may perhaps be ascribed to the fact that this is a solution of a steady-state equation and hence may require a source term; while the physical problem appears to be a quasi-steady-state situation in which the probability density near the origin would be very gradually depleted as the wave function extends further into expanding space.

The two separate terms appearing in the general solution in Eqn. (10) above appear to describe standing wave type solutions, much as do sine or cosine terms in the wave function of a free particle in ordinary static space. It would seem that a complex combination of these terms will provide a suitable description of a propagating wave similar to the manner in which a complex combination of sine and cosine terms form a complex exponential wave function $e^{2\pi ikx}$ to describe a propagating wave for a free particle (or as Hankel functions, which are complex combinations of Bessel functions, describe propagating cylindrical wave solutions to the Bessel equation in cylindrical coordinates).

These exact solutions to the governing equation show that the quantum wave functions are concentrated and localized by this governing equation, and that the measures of the radii of the regions of high wave amplitude or concentration of probability associated with this localization are in good approximate agreement with the rough values developed already. These values are generally are roughly comparable to the quantity $(h/\pi m H_o)^{1/2}$. We will take the quantity:

$$r_o = (h/\pi m H_o)^{1/2} \qquad (11)$$

as a measure of the radial size of the concentrated region of high probability associated with a quantum object of mass m in a space subject to Hubble expansion.

**Implications for classical behavior**

These results strongly suggest that a quantum object in an expanding universe similar to ours will actually undergo a concentration or localization of its wave function that is caused by the expansion of space (counterintuitive as this may seem), even as the further reaches of its wave function are in principle extending flow into the enlarging universe.

Eqn. (11) appears to show that the radial distance over which the wave function is large depends inversely on the square root of the mass of the quantum object. Accordingly, these results indicate that massive objects will have wave functions characterized by small-sized regions of high probability density, while objects of lower mass will have more extended regions in which they are likely to be found.

**Quantum versus classical behavior in an expanding universe: a criterion**

Can this phenomenon of more stringent localization of high mass objects in an expanding universe be related to the observed fact that larger objects including ordinary human-sized objects behave classically, while very low mass objects behave quantum mechanically in our world?

In order to address this issue further, we need to consider how we distinguish classically behaved objects from quantum objects. Classically behaved objects exhibit negligible wave behavior as entire objects, while quantum objects exhibit very prominent and pronounced wave behavior as entire objects. Let us consider a semi-classical extended object. If this object is to behave classically, any wave behavior associated with it must be very limited, and essentially invisible to the exterior world, and thus largely confined to a region within the object's interior. On the other hand, if this object is to behave quantum mechanically, one would expect its wave behavior to extend far beyond the limits of its physical size. This observation provides us with a criterion for distinguishing classical from quantum behavior for an extended object. If the large values of the quantum wave function are concentrated into a region considerably smaller than the size of the extended object, we can classify the object as a classically behaved object, at least as an approximation. In contrast, if large values of the quantum wave function associated with the object extend well beyond the physical size of the object, we can regard the object as having manifest quantum behavior. A threshold between largely classical and largely quantum behavior would then be located at the boundary between these behaviors.

We can characterize the physical size of an extended object by a length L. If we use as an estimate for the size of the region of concentration for the quantum wave function, the quantity $r_o$ derived above (or one of the similar estimates), then we can set $L = r_o$ to define a threshold value for the size that separates largely classical from largely quantum behavior. Following this approach, we can use Eqn. (11) to estimate the threshold size $L_T$ for the transition between quantum and classical behavior due to the expansion of the universe as:

$$L_T = (h/\pi m H_o)^{½} \qquad (12)$$

This tells us that any physical object can be expected to behave classically, if its size exceeds the value given above in Eqn. (12). (This is an approximate result, of course.)

It would seem worthwhile to examine numerical values at least briefly. What will the magnitude of the quantity $(h/\pi m H_o)$ be? Clearly, it is smaller for large values of the mass m and larger for small values of the mass. Let's put in some numbers so as to estimate the radial extent of the region of large values for the wave function. For the Hubble expansion coefficient we will use $H_o = 2.3 \times 10^{-18}$ per second and for the Planck constant we will use the value $h = 6.63 \times 10^{-34}$ joule-second. Then the quantity $(h/\pi H_o) = 0.918 \times 10^{-16}$ in mks or SI units, so that $(h/\pi H_o)^{½} = 0.958 \times 10^{-8}$ in SI units.

If we consider the case of an electron of mass 9.11 x $10^{-31}$ kilogram, we find that $(h/\pi m H_o)$ = 2.0 x $10^{15}$ in SI units, so the associated approximate threshold value for the radial length of the central region of high wave amplitude for a free electron would be given approximately by $r_o \approx 10^7$ meters. Thus, for a free electron, these approximations would suggest that the wave function would be essentially constant over a radial distance of the order of ten thousand kilometers, so that for all practical purposes, the wave function of a free electron in the expanding universe would behave similarly to the wave function of a free electron at rest in ordinary static Euclidean space, and not be affected by the Hubble expansion even over tens of thousands of kilometers.

If, on the other hand, we consider an object of more familiar size, of mass 1 gram, then we would find a threshold radius of 3 x $10^{-7}$ meter. This tells us that with these approximations, the center of mass of a 1 gram mass would be localized to within roughly a fraction of a micron by its presence in a space expanding at the observed Hubble rate.

Such a threshold for a transition from quantum to classical behavior can also be expressed in terms of a threshold moment of inertia.[3] Using these results, we can introduce a rough estimate for a threshold moment of inertia associated with the present calculation:

$$I_T \approx m L_T^2 \approx h/\pi H_o \qquad (13)$$

Objects with moments of inertia larger than this threshold amount (approximately $10^{-16}$ kg·$m^2$) would be expected to behave in a largely classical manner based on the response of their wave functions to Hubble expansion alone, while objects with smaller moments of inertia could retain quantum behavior unless brought into classicality by other mechanisms, such as local quantum decoherence.[4,5,6]

**Comparison with results of other studies**

In the present paper, we are following up on earlier studies that have suggested that certain large-scale properties of the universe may have a role in affecting the transition between quantum behavior and classical behavior of objects.[3,7-11] These studies appear to have shown that certain properties of the universe related to its expansion or temporal duration that can be expressed in terms of the Hubble constant or similar parameters can have a role in effecting the quantum-classical transition. These studies led to the conclusion that sufficiently large objects must behave classically, and the studies individually introduce threshold criteria for "sufficiently large". Interestingly enough, these earlier papers that were variously based on different physical arguments have come up with roughly similar criteria for a threshold between quantum and classical behavior. Arguments based on the Heisenberg uncertainty principle in the presence of Hubble expansion; random motion in the context of the stochastic quantum theory; wave packet dispersion over time; and constraints on wave packet behavior due to the finite temporal duration of the universe have all led to closely related criteria for a critical or threshold size separating quantum behavior from classical behavior.[7-10]

The present results are in approximate agreement with these earlier studies and also seem to be in reasonable agreement with observations, in that objects with moments of inertia larger than values comparable to the threshold moment of inertia just presented in Eqn. (13) are without exception observed to behave in a classical manner.[3] It now appears that the response of quantum wave functions to the Hubble expansion can account for this phenomenon.

**Discussion and summary**

In this paper, we have used an idealized model of the expanding universe, in which the expansion is regarded as taking place at every location in the universe. It will be recalled that a unique coordinate system has been used, a radial coordinate system centered on the classical object, and from this center the expansion of the universe enters the problem. Accordingly, while this type of analysis could be applied to any position in the universe, this approach is applicable in its present form only to the particular central point chosen and the quantum object associated with it. This is an issue that would seem to deserve further attention. Another issue of possible concern is the validity of introducing the Hubble expansion velocities into the Schrödinger equation in the manner followed.

Could this effect of the localization of physical objects as a consequence of Hubble expansion be distinguished experimentally from the imposition of classical behavior by other causes? Perhaps observation of a kind of shell structure of the probability density of the center of mass of objects associated with the characteristics of Bessel functions in their wave functions might present a new and different way of distinguishing this effect from other effects that could cause quantum objects to behave in a classical manner.

To summarize, we have seen that in an expanding space, the wave function for an object at rest does not extend uniformly throughout all of space, but rather is concentrated and localized. These results suggest that for such a region of localized large amplitude for the quantum wave function, a radius of approximately $(h/\pi m H_o)^{1/2}$ provides a reasonable rough estimate for the size of the region of localization that characterizes a quantum object existing in a space subject to Hubble expansion. This analysis appears to support the idea that sufficiently large objects in an expanding universe may be forced into classical behavior by cosmological expansion. Thus, this analysis indicates that classical behavior in the case of large objects may simply derive from the expansion of space. That would seem to be a profound conclusion that deserves serious consideration by the scientific community.

This has been an extremely simplified approach to a very complicated and difficult problem. However, we present these results in the hope that they may be of sufficient interest that others will address the question more skillfully and more thoroughly.

**Acknowledgments**

The author wishes to thank Matthew Rodman of Nova Scotia and the S.O.S. Mathematics CyberBoard for assistance in dealing with and identifying solutions for relevant differential equations.

**Appendix A: Cosmological expansion**

We have assumed the existence of an idealized uniform cosmological expansion throughout all of space.[1] While Hubble expansion has been observationally confirmed at large distances, it has not been confirmed at small distances; however, the possibility of effects associated with an underlying expansion of space observable at distances smaller than cosmological distances should not be overlooked, and there has been some interest in whether the Hubble expansion affects phenomena at smaller distances.[12,13]

Hubble expansion is customarily discussed in conjunction with cosmic distances, and especially for gravitational systems not bound to each other. There is evidence that Hubble expansion may not be manifested in the motions of stellar objects in large gravitationally bound systems such as galaxies (a suppression identified as in conformity with the virial theorem). And at smaller scales matter has clumped together under the influence of gravitational attraction and these clumps do not individually appear to expand, though they continue to recede from one another.[1] The question has been discussed as to whether there might exist any cut-off distance such that systems below that scale do not participate in the Hubble expansion, and, if so, what could possibly determine such a maximum scale.[14,15] The cosmological metric alone does not dictate a scale for expansion, and it seems difficult to justify any particular scale for the onset of expansion or the 'shielding' of systems smaller than that scale from cosmic expansion.[14] Evidence for Hubble expansion has been sought at distances less than cosmic distances, and some evidence for residual Hubble expansion effects has been reported from lunar laser ranging measurements.[16,17]

But what about even smaller scales, such as those relevant to present concerns? The effect is actually registered over the very small distances of the wavelengths of light, although this is sometimes regarded as an imprint of the expansion at larger scales.[15] It has been argued that there is only one space-time and that therefore all physical systems, large and small, must feel the effects of this cosmic expansion in one way or another, and it has been suggested that in principle Hubble expansion could be present at the smallest practical scale and observable in principle.[14,15] For present purposes, we will make the assumption that Hubble expansion is present at the smallest scales and that every point in space is a locus of expansion. While there may be some question as to whether Hubble expansion does in fact takes place at small distances, it would appear to be worth examining the possible consequences of local cosmic expansion on physical objects.[7]

**Appendix B: Development and use of a wave equation to govern quantum wave functions in an expanding space**

**Developing a wave equation**

It would seem to be worthwhile to direct some effort towards developing a wave equation appropriate to the description of a quantum object in an expanding universe. We will not attempt the use of a relativistic wave equation, but rather simply explore the possible application of the Schrödinger equation in order to examine possible wave characteristics in the region of non-relativistic behavior. Since we are concerned with quasistatic or slowly varying solutions, an approximate solution will be sought with the use of the time-independent Schrödinger equation. Neither the Schrödinger equation for a free particle nor the Schrödinger equation for a particle in a potential would be entirely appropriate in this context. However, the Schrödinger equation for a particle in a potential, which involves the difference between the total energy and the potential energy, can be adapted by being reexpressed in terms of the kinetic energy and thus in terms of the magnitude of the velocity.

In three dimensions, under spherically symmetric conditions, the time-independent Schrödinger wave equation can be written as the product of radial and angular wave functions, with the radial wave equation being given by:[2]

$$(1/r^2)(d/dr)(r^2 d\psi(r)/dr) + (8\pi^2 m/h^2)[E - V(r) - (h^2/8\pi^2 m)\{\ell(\ell+1)/r^2\}]\psi(r) = 0 \quad (B1)$$

Here, $\ell$ is the orbital angular momentum quantum number characterizing the angular wave function. If we limit our attention to the case of wave functions with zero angular momentum, the last term drops out. Limiting our attention to zero angular momentum conditions, the preceding equation, Eqn. (B1), would then become:

$$(1/r^2)(d/dr)(r^2 d\psi(r)/dr) + (8\pi^2 m/h^2)[E - V(r)]\psi(r) = 0 \quad (B2)$$

The difference between the total energy and the potential energy is the kinetic energy K, which is given by $E - V = K = \frac{1}{2}mv^2$. Let us reexpress the Schrödinger equation in terms of a local velocity function defined by:

$$\frac{1}{2} m v(r)^2 = E - V(r) \quad (B3)$$

The governing equation then becomes:

$$(1/r^2)(d/dr)(r^2 d\psi(r)/dr) + (4\pi^2 m^2/h^2) v(r)^2 \psi(r) = 0 \quad (B4)$$

But, since in this problem we are considering an object that is at rest at the origin of coordinates, and the only velocity present is the expansion velocity which is proportional to the radial distance from the origin:

$$v(r) = H_o r \quad (B5)$$

The governing differential equation becomes:

$$(1/r^2)(d/dr)(r^2 d\psi(r)/dr) + (2\pi m H_o/h)^2 r^2 \psi(r) = 0 \qquad (B6)$$

Let us pause to examine some features of this governing equation.

First off, if the mass is vanishingly small, the second term will be very small, and in the limit, we would be left with the first term only. Thus, we would be left with the ordinary Schrödinger equation for a free particle with zero kinetic energy, that is, an ordinary free particle at rest in a static universe. So, in the limit of sufficiently small masses, objects governed by this equation would tend to behave as ordinary quantum objects at rest.

Secondly, when the mass m is sufficiently large, the second term will become very large. In the limit of large mass, both terms can remain finite only if the product ($r^2\psi(r)$) is zero. This would take place if r is zero and/or if $\psi(r)$ is zero. Thus, in the limit of high masses, the wave function can be non-zero only near the origin of coordinates, or in the limit, at the origin of coordinates. So this result tells us that it is an intrinsic feature of this governing equation for quantum objects in an expanding space that it will cause the wave functions to concentrate so that objects of sufficiently large mass will become strongly localized near the origin of coordinates.

Thus, it appears that this governing equation will cause massive objects to be localized, while particles with small masses will behave as delocalized quantum objects.

Eqn. (B6) can be reexpressed as:

$$d^2\psi(r)/dr^2 + (2/r)d\psi/dr + (2\pi m H_o/h)^2 r^2 \psi(r) = 0 \qquad (B7)$$

It would appear that solutions to this differential equation might be developed in a manner somewhat similar to the conventional approach used for the harmonic oscillator, or perhaps more generally by the use of the method of variation of parameters or similar techniques.[2,18,19]

We can also reexpress Eqn. (B6) in the form:

$$r\, d^2\psi(r)/dr^2 + 2\, d\psi/dr + (2\pi m H_o/h)^2 r^3 \psi(r) = 0 \qquad (B8)$$

Eqn. (B8) can be identified as a Bessel-like differential equation, with solutions involving Bessel functions.[19,20]

**Solutions to the wave equation**

The solution to this governing equation given in Eqn. (B8) can be expressed in terms of a linear combination of Bessel functions of fractional order:[20,21]

$$\psi = r^{-1/2} [C_1 J_{1/4} (\pi m H_o r^2/h) + C_2 Y_{1/4} (\pi m H_o r^2/h)] \tag{B9}$$

Here $C_1$ and $C_2$ are arbitrary constants. $J_{1/4}$ is a Bessel function of the first kind of fractional order ¼ and $Y_{1/4}$ is a Bessel function of the second kind (also called a Neumann function) of fractional order ¼.[21]

How do these functions involved in the solution to the governing equation behave? Bessel functions of the first kind are typically smooth oscillatory functions that look qualitatively like sine or cosine functions that decrease in amplitude with increasing value of the argument. Bessel functions of the second kind are typically infinite at the origin but otherwise behave qualitatively in a somewhat similar fashion to Bessel functions of the first kind.

To simplify the numerical treatment, we will introduce a new variable, and write $x = \pi m H_o r^2/h$. Then, $r = (h/\pi m H_o)^{1/2} x^{1/2}$ and $r^{-1/2} = x^{-1/4} (\pi m H_o/h)^{1/4}$. Eqn. (B9) can then be written in the form:

$$\psi = (\pi m H_o/h)^{1/4} x^{-1/4} [C_1 J_{1/4}(x) + C_2 Y_{1/4}(x)] \tag{B10}$$

From tabulations of Bessel functions we can obtain numerical evaluations of these Bessel functions.[22] These tables show that $J_{1/4}(x)$ is an oscillating function whose amplitude decreases with increasing x, and they also show that $J_{1/4}(x)$ is zero at the origin, while the term $x^{-1/4} J_{1/4}$ is finite at the origin.

Let us now look at the second term of the solution in Eqn. (B9) or (B10).

Bessel functions of the second kind having non-integer order can be evaluated in terms of Bessel functions of the first kind using the equation:[23]

$$Y_\alpha(x) = [J_\alpha(x) \cos(\alpha\pi) - J_{-\alpha}(x)]/\sin(\alpha\pi) \tag{B11}$$

For the case of Bessel functions of order ¼, Eqn. (B11) indicates that:

$$Y_{1/4}(x) = J_{1/4}(x) - \sqrt{2} J_{-1/4}(x) \tag{B12}$$

Calculated values for both $J_{1/4}(x)$ and $J_{-1/4}(x)$ are tabulated.[22] Both $J_{1/4}(x)$ and $J_{-1/4}(x)$ exhibit oscillatory behavior of decreasing amplitude as a function of x, and they appear to be to some extent out of phase with each other, so that their combination in Eqn. (B12) would seem capable of damping the oscillatory behavior of these functions even further. However, since $J_{-1/4}(x)$ is infinite at $x = 0$ while $J_{1/4}(x)$ is zero at $x = 0$, Eqn. (B12) tells us $Y_{1/4}(x)$ will go to (negative) infinity at $x = 0$. Furthermore, the product $x^{-1/4} Y_{1/4}$ would be expected to have a finite first term (since $x^{-1/4} J_{1/4}$ is finite at $x = 0$), but because of the second term it would also be expected to become infinite at $x = 0$.

We will now concentrate our attention on the first term in the solutions given in Eqn. (B9) or (B10), which remains finite.

We saw that the general solution to the equation for a quantum wave function in an expanding space could be expressed as the product of a linear combination of Bessel functions of fractional order with a factor $r^{-1/2}$. We now see that each of the corresponding basic solutions exhibits localization in the sense that the wave functions exhibit their largest amplitudes at generally small values of the radial distance, while at larger radial distances the magnitudes of the oscillations decrease. As a consequence, the probability of finding the quantum object at a large distance from the origin is small in these solutions. Thus, it would appear that all such solutions to the governing equation for a quantum object in an expanding space will impose appreciable localization on the quantum object.

We can obtain an estimate of the extent of the quantum wave from any of these solutions by looking at the first zero in the radial behavior beyond the origin of coordinates. We also take into account the decrease with r of the functions involved.

Thus, examining tabulated values for $J_{1/4}(x)$, we see that it is zero at the origin, reaches a maximum of approximately 0.77 near x = 0.77, drops to zero again at x = 2.78, goes to a negative maximum of about - 0.385 at a value of x of about 4.22, and continues to larger values of x with oscillations of decreasing magnitude.

To examine the behavior of the first term in the solution in Eqn. (B10), let us look at the behavior of $x^{-1/4} J_{1/4}(x)$. This function takes on a tabulated value of approximately 0.928 at x = 0 (corresponding to r = 0). It drops off to half-height near x = 1.7 (corresponding to r equal to about $1.3(h/\pi m H_o)^{1/2}$), and drops further to its first zero in the vicinity of x = 2.78 (corresponding to r = $1.67(h/\pi m H_o)^{1/2}$). The function then goes negative, and at a value of x of about 4.2 (corresponding to a value of r of approximately $2(h/\pi m H_o)^{1/2}$), it peaks at about - 0.27, or about 30% of the maximum value at x = 0. It then rises to another zero at around x = 5.9 (or r = $2.4(h/\pi m H_o)^{1/2}$) and then goes up to some positive values and continues the decaying oscillation at increasing values of x or r. The probability density varies as the square of the wave function amplitude, so the maximum probability density beyond the first zero would be only about 1/10 or less of the probability density at the origin, so it would seem clear that an approximate measure of the size of the region of localization of the quantum wave function would correspond to a radial value comparable in magnitude to the value of $(h/\pi m H_o)^{1/2}$, or a few times this value, at most.

Similarly, we can examine the behavior of the second term in the solution of the governing equation in Eqn. (B10) by making some simple calculations using tabulated values of Bessel functions. We can evaluate of $x^{-1/4} Y_{1/4}(x)$ for small and moderate values of x. This function is infinite at x = 0 (corresponding to r = 0) and drops rapidly to its first zero at roughly x = 1.2, corresponding approximately to r = $1.1 (h/\pi m H_o)^{1/2}$. The value $(h/\pi m H_o)^{1/2}$ can therefore serve as an estimate for the size of the region of localization associated with this singular solution as well.

While solutions that become infinite are generally undesirable, the fact is that the proposed governing equation describes a steady-state situation, while in the physical problem we are looking at a quasi-steady-state situation. In the quasi-steady-state situation, no source term at r = 0 would be needed; rather, the near-zero region would become depleted with time. Therefore, it would appear that we can treat the singular solution as one that would become large but finite in a more realistic treatment.

While the two independent parts of the solution to the governing equation appear to represent static solutions with no net flow, any combination of these is also a solution, and it would seem that suitable combinations could be developed to represent actual flow in terms of running waves analogously to the way that free particle solutions of the ordinary Schrödinger equation that represent standing waves, $\sin(2\pi kx)$ and $\cos(2\pi kx)$, can be combined in the form $e^{2\pi ikx}$ to form plane running waves. These complex combinations of Bessel functions would also be analogous to the Hankel functions ($H^{(1)}_\alpha (x) = J_\alpha(x) + iY_\alpha(x)$ and $H^{(2)}_\alpha (x) = J_\alpha(x) - iY_\alpha(x)$) which are two linearly independent solutions to Bessel's equation (and are otherwise known as Bessel functions of the third kind) that are used to express outward- and inward-propagating cylindrical wave solutions of the cylindrical wave equation.[23]

So we see that in an expanding space, the wave function for an object at rest does not extend uniformly throughout space, but rather is concentrated within the neighborhood of the associated semiclassical object. These results suggest that a region of concentration or localization of wave amplitude of radius of approximately $(h/\pi m H_o)^{1/2}$ provides a reasonable rough estimate for the size of the wave that characterizes a quantum object existing in a space subject to Hubble expansion.

**References**

1. Wikipedia, "Metric Expansion of Space," http://en.wikipedia.org/wiki/Metric_expansion_of_space (accessed 14 November 2009).

2. David B. Beard, *Quantum Mechanics*, Allyn & Bacon, Boston, 1963.

3. C. L. Herzenberg,"The quantum-classical boundary and the moments of inertia of physical objects," http://arxiv.org/abs/0908.1760 (2009).

4. Erich Joos, "Decoherence and the transition from quantum physics to classical physics," in *Entangled World: The Fascination of Quantum Information and Computation*, Jürgen Audretsch (Ed.), Wiley-VCH Verlag, Weinhem, 2002.

5. Erich Joos, "Decoherence," http://www.decoherence.de/ (accessed 13 July 2009).

6. Wikipedia, "Decoherence," http://en.wikipedia.org/wiki/Quantum_decoherence (accessed 13 September 2009).

7. C. L. Herzenberg, "Becoming Classical: A Possible Cosmological Influence on the Quantum-Classical Transition," *Physics Essays* **19**, 4, 1-4 (2006).

8. C. L. Herzenberg, "The Role of Hubble Time in the Quantum-Classical Transition," Physics Essays **20**, 1, 142-147 (2007a).

9. C. L. Herzenberg, "Why our human-sized world behaves classically, not quantum-mechanically: A popular non-technical exposition of a new idea," http://arxiv.org/abs/physics/0701155 (2007b).

10. C. L. Herzenberg, "The Quantum-Classical Transition and Wave Packet Dispersion," arXiv:0706.1467 http://arxiv.org/abs/0706.1467 (2007c).

11. C. L. Herzenberg, "How the classical world got its localization: An elementary account of how the age of the universe may be implicated in the quantum-classical transition," http://vixra.org/abs/0909.0038 (September 2009).

12. W.B. Bonnor, Mon. Not. R. Astron. Soc. **282**, 1467 (1996).

13. A. Dominguez and J. Gaite, Europhys. Lett, **55** (4), 4558 (2001).

14. J. L. Anderson, Phys. Rev. Lett. **75**, 3602 (1995).

15. F.I. Cooperstock, V. Faraoin, and D.N. Vollick, Astrophys. J. **503**, 61 (1998).

16. Yu.V. Dumin, arXiv:astro-ph/0302008, 28 February 2003.

17. Yu.V. Dumin, arXiv:astro-ph/0203151, 11 March 2002.

18. H. T. Flint, *Wave Mechanics*, Methuen and Co. Ltd., London, 1967.

19. S.O.S. Mathematics CyberBoard, http://www.sosmath.com/CBB/index.php (accessed 18 November 2009).

20. EqWorld, http://eqworld.ipmnet.ru/en/solutions/ode/ode0209.pdf (accessed 18 November 2009)

21. Andrei D. Polyanin and Valentin F. Zaitsev, *Handbook of Exact Solutions for Ordinary Differential Equations*, CRC Press, Boca Raton FL, 1995.

22. U.S. National Bureau of Standards Computation Laboratory, *Tables of Bessel functions of fractional order*, Vols. 1 and 2, Columbia University Press, New York, 1948-1949.

23. Wikipedia, "Bessel function," http://en.wikipedia.org/wiki/Bessel_function (accessed 20 November 2009).

qwavfcnexpandingunivfin.doc
6 December 2009 draft